\pdfoutput=1
\documentclass[11pt]{article}

\usepackage[affil-it]{authblk} % Package for author affiliation
\usepackage[authoryear]{natbib}
\usepackage{gensymb}
\usepackage{amsmath}
\usepackage{amssymb}
\usepackage{graphicx}
\usepackage{float}
\usepackage{booktabs}
\usepackage[a4paper, top=1in,bottom=1in,left=1in,right=1in,marginparwidth=0cm]{geometry}
\usepackage[title]{appendix}
\usepackage{pdflscape}
\usepackage{longtable}

\usepackage{adjustbox}
\usepackage{array}
\usepackage{caption}
\usepackage{chngcntr}
\usepackage{mathtools}
\usepackage{lineno}
\usepackage{setspace}
\usepackage{makecell}
\usepackage{hyperref}
\hypersetup{colorlinks=true,allcolors=blue}

\newcolumntype{R}[2]{%
    >{\adjustbox{angle=#1,lap=\width-(#2)}\bgroup}%
    l%
    <{\egroup}%
}
\renewcommand{\eqref}[1]{Eq.\,\ref{#1}}

\title{Negative density-dependent dispersal of the mountain pine beetle in Alberta}
\author[1,*]{Evan C. Johnson}
\author[2,3]{Mark A. Lewis}
\affil[1]{Mathematical and Statistical Sciences; University of Alberta; Edmonton, Alberta, Canada}
\affil[2]{Department of Mathematics and Statistics; University of Victoria; Victoria, British Columbia, Canada}
\affil[3]{Department of Biology; University of Victoria; Victoria, British Columbia, Canada}
\affil[*]{Corresponding author: Evan Johnson, ecjohns1@ualberta.ca}

\date{} % Date of the document, you can set a specific date by replacing \today with the desired date

\begin{document}

\maketitle % Generates the title, author, and date

\newpage

\tableofcontents % Generates the table of contents

\newpage 

\section*{Abstract} 

Understanding the mountain pine beetle's dispersal patterns is critical for evaluating its threat to Canada's boreal forests. It is generally believed that higher beetle densities lead to increased long-distance dispersal due to aggregation pheromones becoming repellent at high densities, causing beetles to seek areas with less competition. However, using helicopter surveys of infested trees, along with statistical models, we find no evidence supporting a positive relationship. Instead, we observe a weak negative association between population density and dispersal at all spatial scales. A possible explanation is that at low population densities, beetles cannot successfully attack healthy trees and must travel farther to find weakened hosts. Even so, the influence of beetle density on dispersal is minor compared to the spatiotemporal variation in the overall (density-independent) scale of dispersal, as revealed by our models. This variation accounts for the MPB's erratic range expansion across western Alberta, which varied from 20 km to 220 km annually.

\newpage

\section{Introduction} \label{Introduction}

Bark beetle outbreaks in North America are worsening due to climate change. Warmer winters reduce larval mortality \citep{carroll2006impacts}, whereas drought conditions weaken host trees' defenses \citep{alfaro2009historical, creeden2014climate}. Consequently, several species of bark beetles --- most notably the mountain pine beetle --- have had substantial socioeconomic impacts, harming timber-based industries and transforming Canada's coniferous forests into a net carbon source \citep{corbett2016economic, abbott2009mountain, dhar2016consequences, kurz2008mountain}. 

Irruptive bark beetle species share similar natural histories (\citealp{raffa2015natural}; \citealp{safranyik2006biology}). These insects spend the majority of their life cycles as larvae beneath the bark of host trees. The larvae feed on the tree's phloem or cambium during warmer periods. Generation time varies from two years (i.e., semivoltine) to multiple generations per year (i.e., multivoltine), with overwintering beetles utilizing cryoprotectants to avoid freeze-thaw damage. Adults emerge briefly only in spring or summer for dispersal. Using aggregation pheromones and synchronized emergence (triggered by temperature-dependent developmental thresholds; \citealp{logan2001ghost}), the beetles launch coordinated attacks on trees. These \textit{mass attacks} function to overwhelm a tree's defensive capability, which primarily involves exuding toxic resin. Once the tree's resin supply is depleted, the beetles can successfully bore into their host. 

To predict the spatial spread of irruptive bark beetles, we must understand their complex dispersal behavior, which is thought to be influenced by (among other things) wind, temperature, pheromones, host-tree volatiles, microclimate, topography, and individual beetle physiology \citep{jackson2008radar, byers2004chemical, gaylord2014influence, franklin1999flight, jones2019influence,  amman1998silvicultural, de2011incoming, jones2021effect}. Despite extensive research, identifying the most important predictors of dispersal distance remains challenging. Even obtaining accurate in situ estimates of average dispersal distance is difficult. Tracking individual beetles is impossible, and while mark-recapture experiments have been conducted \citep{zumr_dispersal_1992, turchin_quantifying_1993, zolubas_recapture_1995, werner_dispersal_1997, dodds_sampling_2002}, but the use of pheromone traps for recapture may negatively bias dispersal distance estimates. Modeling approaches present their own challenges, as assumptions can significantly alter conclusions. For example, model-based estimates of typical mountain pine beetle dispersal distance range from 10 meters to 18 kilometers \citep{johnson2024stratified}.

Understanding the dispersal of the mountain pine beetle (MPB; \textit{Dendroctonus ponderosae} Hopkins) is particularly important, given that MPB has recently expanded its range from British Columbia into Alberta and now threatens boreal forests stretching eastward to Quebec \citep{safranyik2010potential, cooke2017predicting, bleiker2019risk}. MPB is thought to exhibit three quasi-discrete dispersal behaviors, leading to a continuum of dispersal distances from zero meters to hundreds of kilometers. First, the majority of beetles engage in short-range dispersal behavior. These beetles are sensitive to host-tree cues and aggregation pheromones and thus rarely disperse beyond 100 meters \citep{safranyik1992dispersal, robertson2009spatial}. Second, a minority of beetles, primarily females in good energetic condition characterized by their large body size and high lipid reserves \citep{shegelski2019morphological}, can disperse up to several kilometers under the canopy \citep{borden1993uncertain, robertson2007mountain, evenden2014factors}. This behavior is thought to be an evolutionary stable bet-hedging strategy: although most pioneers fail, the successful few that find high-quality host trees achieve high fitness \citep{raffa2001mixed, kautz2016dispersal}. Third, a small fraction of beetles fly above the canopy, where atmospheric winds may carry them up to 300 kilometers \citep{furniss1972scolytids, Hiratsuka1982, Cerezke1989, jackson2008radar}.  A mark-recapture study estimated that 2.5 \% of the MPB adults fly above the canopy \citep{safranyik1992dispersal}. However, it is generally thought that MPB’s propensity for this behavior could be much higher, especially in heavily infested stands.

The relationship between MPB population density and long-distance dispersal may have implications for forest management. During the so-called ``hyperepidemic'' of the 2000s, MPB jumped over the Rocky Mountains, a former geographic barrier, and spread to the central longitudes of Alberta in just 4 short years \citep{cooke2017predicting}. Subsequent eastward spread was slow, which can be attributed to two main factors. First, eastern Alberta has lower pine volumes compared to the west \citep{bleiker2019risk, safranyik2010potential}. Secondly, Jack pine, an evolutionary novel host tree that inhabits eastern Alberta, may be less suitable for the beetle \citep{bleiker2023suitability, srivastava2023dynamic}. The primary risk of range expansion is now thought to be long-distance dispersal from high-volume pine stands in central Alberta to high-volume stands in western Saskatchewan \citep{bleiker2019risk, hodge2017strategic}.

It is generally assumed that long-distance dispersal is more common at higher MPB densities \citep{carroll2003bionomics, de2011incoming, bone2014modeling, bleiker2019risk}, though there is little direct evidence. There are, however, several circumstantial reasons to expect such a relationship. 1) In trees experiencing heavy mass-attacks, MPB produces the anti-aggregation pheromones verbenone and frontalin \citep{ryker1982frontalin, borden1998volatiles}. Additionally, the pheromone \textit{ex}-brevocomin, which is normally attractive to female MPB at low concentrations, becomes repellent in sufficiently high concentrations \citep{rudinsky1974antiaggregative}. These semiochemicals ``tell'' beetles to escape a highly competitive environment, which may manifest as beetles flying above the canopy. 2) In the epidemic phase of an outbreak, there are sufficient beetles to attack and kill large pine trees. These trees have thick phloem, leading to beetles with a good energetic condition \citep{graf2012association}, which is positively associated with flight distance \citep{shegelski2019morphological}. 3) The two years in which many beetles dispersed from BC to Alberta (2006 \& 2009; \citealp{carroll2017assessing}) occurred during the peak years of the outbreak in British Columbia. On the other hand, there are reasons to expect a negative relationship between MPB density and typical dispersal distance. Perhaps beetles have to go farther at low beetle densities, since they do not have the numbers to attack well-defended trees; instead they must find recently compromised trees \citep{bleiker2014characterisation}.

There is mixed evidence for a statistical relationship between population density and long-distance dispersal. First, \citet{chen2011mountain} found a negative correlation between long-distance dispersal and the total number of infested areas within British Columbia. In contrast, \citet{powell2014phenology} developed a model implying that typical dispersal distances decrease (from kilometers to meters per day) as host tree density decreases. Similarly, \citet{goodsman2017positive} fit a statistical model showing lower beetle recruitment when the ratio of beetles to susceptible trees was high. The latter two studies indirectly support a positive relationship between density and dispersal, as beetle density is likely to be high during the first year when host tree density becomes very low. These contrasting findings highlight the need for further research.

Here, we provide empirical evidence regarding the relationship between MPB density and dispersal distance using high-quality data and statistical models. Specifically, we utilized helicopter surveys, the most accurate remote sensing method for bark beetle infestations \citep[Table 2.]{coggins2008linking}. Our analysis employed simple models with a proven track record of characterizing MPB dispersal. While we anticipated a positive relationship, our findings revealed a small negative relationship. These results challenge existing assumptions and contribute to the ongoing dialogue about MPB dynamics and their implications for forest management.

\section{Methods} \label{Methods}

\subsection*{Data}

Alberta Forestry and Agriculture's Mountain Pine Beetle (MPB) management program collects data annually through helicopter surveys (termed \textit{Heli-GPS surveys}) and ground surveys. Around September, helicopters fly a  ``lawnmower'' search pattern over large portions of Alberta, recording the locations of \textit{red-topped trees}. These trees, characterized by rust-red needles, were first attacked the year prior. Because MPB is univoltine, the progeny from the red-topped trees have already emerged and are likely infesting nearby \textit{green-attack trees}, trees which have been successfully attacked but are not yet displaying red needles. Field crews are sent to the location of the red-topped tree to manually search for green-attack trees, using clues such as boring dust and pitch tubes. Infested trees are the ``sanitized'', typically by cutting and burning. The quality assurance protocols ensure that at least 95\% of the Heli-GPS points are within $\pm$ 30 meters of the true location, and that 90\% of the Heli-GPS points record the number of infested trees within a narrow error tolerance (the range changes depending on the true number of trees; \citealp[p. 21]{government2016mountain}). A previous study found that infestation-severity classifications were 92\% accurate with Heli-GPS surveys \citep{nelson2006large}.

\begin{figure}[H]
\centering
\includegraphics[scale = 1]{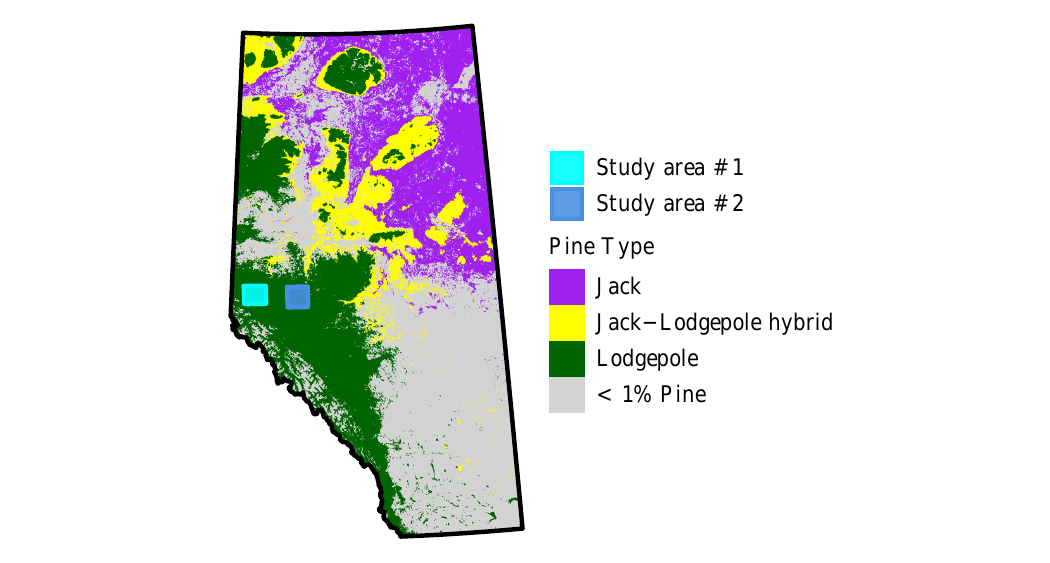}
\caption{Location of the two study areas. Pine species data comes from \citet{cullingham2012characterizing}. The grey pixels are areas where pines constitute less than 1\% of total live above-ground biomass; data from \citet{beaudoin2014mapping}.}
\label{fig:DD_study_area_smallest_both}
\end{figure}

We focused on two patches in western Alberta, each approximately 2500 $\text{km}^2$ and separated by 50 km (edge-to-edge; Fig. \ref{fig:DD_study_area_smallest_both}). These patches were chosen because they are high-quality MPB habitats, with high-biomass lodgepole pine forest, a moderate thermal regime, and relatively low topographical complexity. We focused on modeling the years 2010--2020, excluding earlier years due to the confounding influence of large beetle migrations from British Columbia in 2006 and 2009 \citep{carroll2017assessing}.

The raw Heli-GPS and ground survey data were rasterized into 30 x 30 pixels,  a resolution that roughly matches the Heli-GPS error margin. The number of infested trees in pixel $x$ and in year $t$ is computed as
\begin{equation}
I_t(x) = c_{t}(x) + r_{t+1}(x),
\end{equation}
where $c_{t}(x)$ is the number of green-attack trees that were located and sanitized in the focal year, and $r_{t+1}(x)$ is the number of red-topped trees observed in the following year. While $I_t(x)$ is a suitable model output (i.e., a response variable), it is not a suitable model input --- it contains controlled trees that do not produce any beetle progeny. Therefore, we define a modified input variable, $I_{t}^{*}(x) = I_t(x) - c_t(x)$.

\subsection*{Model outline}

Our dispersal model considers all possible points of origin for each new infestations. Conceptually, the model operates by predicting the locations of infestations in the focal year (\textit{offspring infestations}) stemming from each infestation in the prior year (\textit{parental infestations}). Then, these individual-level distributions are normalized to create a spatially-explicit probability mass function. 

\begin{figure}[H]
\centering
\makebox[\textwidth]{\includegraphics[scale = 1]{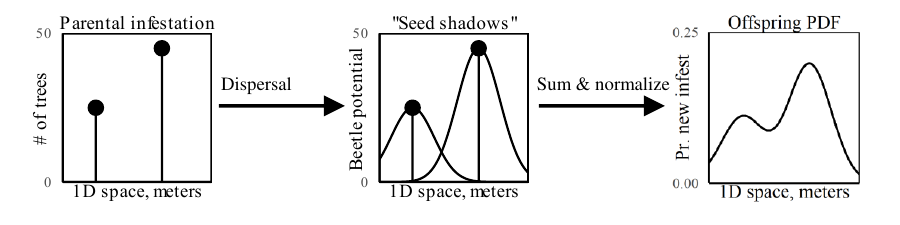}}
\caption{Graphical representation of our redistribution models in one dimension. A dispersal model is applied to each parental infestation, creating surfaces that represent the potential of offspring infestations stemming from parental infestation, termed ``seed shadows'' in the plant-focused dispersal literature. These are summed and normalized to produce a likelihood function for new infestations. Figure originally appears in \citet{johnson2024stratified}.}
\label{fig:model_conceptual_fig}
\end{figure}

Specifically, we utilize a radially symmetric Student's $t$-distribution, a dispersal kernel that has been validated by previous research \citep{johnson2024stratified}. Due to radial symmetry, the kernel can be parameterized as a function of the Euclidean distance between coordinates, $r = dist(x,y)$, where $x$ and $y$ index pixels in 2D space. Although we will use the term ``dispersal kernel'' for its familiarity, it is imprecise in the sense that it implies beetle movement. We are modeling the spatial relationship between infestations in successive years, so the term ``redistribution kernel'' is more apt. We discuss the issues with using infestations as a proxy for beetles further in the \textit{Discussion}. The Student's $t$ dispersal kernel is written as
\begin{equation} \label{eq:student}
D(r) = \frac{(\nu -1) \left(\frac{r^2}{\nu  \rho ^2}+1\right)^{\frac{1}{2} (-\nu -1)}}{2 \pi  \nu  \rho ^2}.
\end{equation}

The likelihood function is computed by convolving parental infestations with the dispersal kernel:
\begin{equation} \label{eq:convolve}
B_t(y) = \sum_{x} I_{t-1}^*(x) D\left(\text{dist}(y, x)\right) \left( \Delta x \right)^2,
\end{equation}
with $\Delta x = 0.03 \text{km}$ as the spatial resolution. The result is \textit{beetle potential} $B_t$, which represents the mean number of offspring infestations received by each pixel; here, the ``mean'' refers to an average of instantiations of a stochastic dispersal process.

The beetle potential is re-scaled to produce a likelihood surface:
\begin{equation}
\pi_t(x) = \frac{B_t(x)}{\sum_y I_{t-1}^*(y)}
\end{equation}

The convolution is performed for all infestations within the approximately 50 x 50 km study area, but the likelihood is only calculated using the offspring infestations in a smaller inset of the study area, defined by a negative 10 km buffer. Without this trick, the model might incorrectly ``think'' that beetles near the edge of the large area came from far away, simply because it can't account for beetles that started just outside the large area.

Each offspring infestation is treated as an i.i.d. event, and thus the log likelihood in year $t$ is
\begin{equation}
\mathcal{L}_t = \sum_{x:I_t(x) > 0} I_t(x) \times \log(\pi_t(x)). 
\end{equation}

\subsection*{Density-dependent dispersal models}

We examine the relationship between beetle density and dispersal distance in two ways. First, we performed a \textit{non-local dispersal analysis}, fitting the dispersal kernel parameters separately to each of the 22 unique combinations of year and study area. Then, we examined the relationship between infestation density and summary statistics of the fitted dispersal kernel (e.g., the median redistribution distance). This approach has the benefit of not assuming the functional form of the relationship between density and dispersal; however, it assumes that the infestation density, averaged over the entire study area, is representative of the beetle density within local neighborhoods.

Individual beetles likely have a sensory neighborhood with a radius of 20--1000 meters. Evidence from small-scale spatial maps and mechanistic pheromone diffusion models suggests that beetles primarily respond to aggregation pheromones that are a maximum of 20--100 meters from their tree of origin (\citealp[Fig. 4--5]{mitchell1991analysis}; \citealp{strohm2013pattern}). However, some beetles are caught in pheromone traps up to 1 km from a point of release, likely due to a combination of diffusive flight and pheromone detection \citep{barclay1998trapping}.

To represent MPB's limited sensory abilities, we also utilized \textit{neighborhood-based dispersal models}. These are hierarchical models in which the dispersal parameters ($\rho$ and $\nu$) are location-specific and depend on the infestation density in a local neighborhood:

\begin{equation} \label{eq:rho}
\rho_{t}(x) = \exp\left[\alpha_0 + \alpha_1 \log\left(\sum_{y \in N(x)} I_{t-1}^*(y) \right)\right],
\end{equation}

\begin{equation} \label{eq:nu}
\nu_{t}(x) = \exp\left[\beta_0 + \beta_1 \log\left(\sum_{y \in N(x)} I_{t-1}^*(y) \right)\right] + 1.
\end{equation}

Here, $N(x)$ is a circular neighborhood around the location $x$. We fit the hierarchical models separately for several fixed neighborhood sizes: 0, 0.003, 0.025, 0.636, 2.545, and 10.179 $\text{km}^2$. These neighborhoods led to similar model performance (Fig. \ref{fig:DD_dens_boxplot}). Consequently, all further analysis was conducted with the 0.025 $\text{km}^2$ neighborhood, which is equivalent to a sensory radius of 90 meters.

The structure of the neighborhood-based models is justified by graphical evidence: the log-log linear relationship between the dispersal parameters ($\rho$ \& $\nu$) and the beetle density emerges from the non-local dispersal analysis (Fig. \ref{fig:density_params} in \textit{Results}). We fit the neighborhood model to each $\text{year} \times \text{site}$ subset of the data to avoid spurious correlations between dispersal distance and beetle density, which may be driven by third variables that vary substantially from year to year, e.g., minimum winter temperature, wind speed during the flight period, and the density of suitable host trees remaining. 

All models were fit with maximum likelihood estimation using the Nelder-Mead optimization algorithm. To monitor for local optima, we repeatedly fit each data subset with random initial parameters and manually checked that the maximum likelihood parameters converged to nearly identical values.  All analyses were performed in R \citep{R2020language}. 

We faced significant computational challenges due to the necessity of brute force convolution, which scales as $O(n^4)$, where $n$ is the linear dimension of our study area. Parameter estimates were unstable with larger pixel sizes, as most dispersal occurred within 100 meters of the natal tree. To reduce the computational burden, we limited dispersal kernels to a 5 km radius from parental trees. Inferences made about dispersal across longer distances are therefore based on extrapolating the shape of the dispersal kernel. Offspring infestations that appeared further than 5 km from any parental infestations were extremely rare; to avoid an infinite log likelihood stemming from these few observations, we added machine epsilon to each location's probability mass and renormalized. Despite these adjustments, computations still required approximately 48 hours on a 40-core computer.

\section{Results} \label{Results}

\begin{figure}[H]
\centering
\includegraphics[scale = 1]{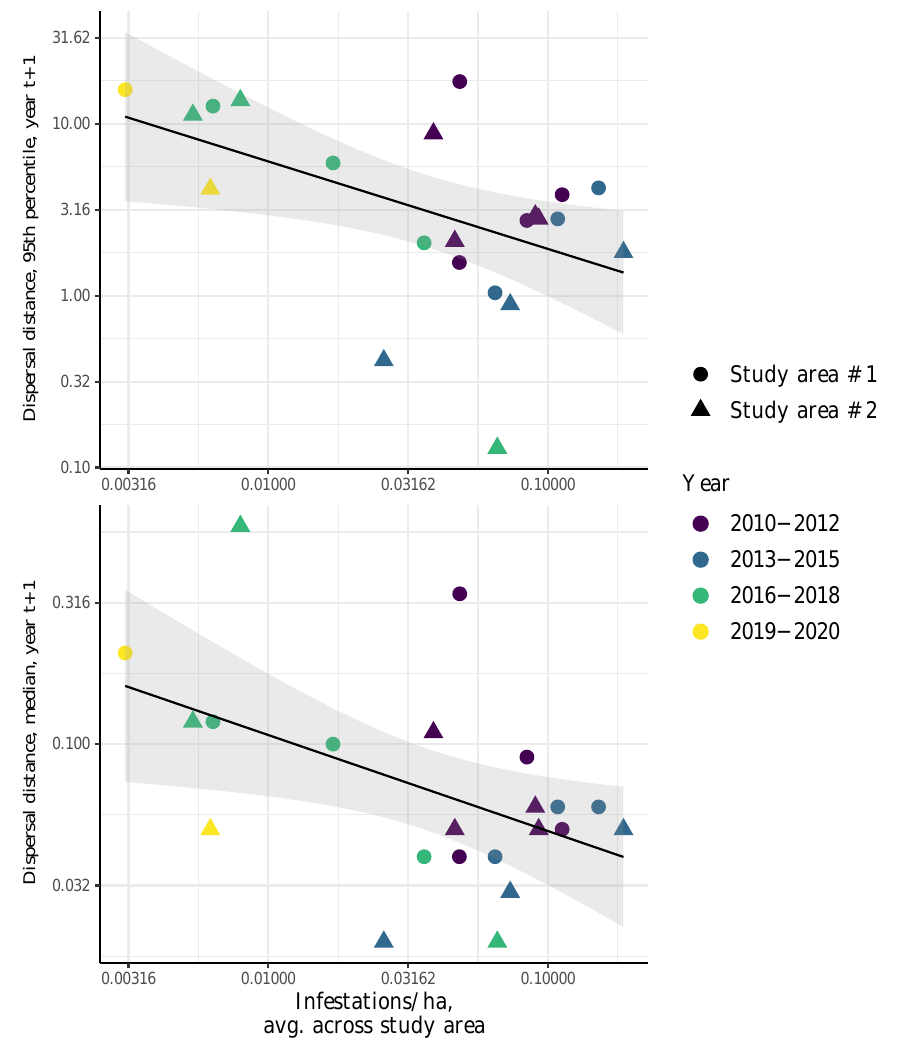}
\caption{There is a negative association between global (across a study area) infestation density and dispersal distance. Each point represents a unique combination of study year and study area.}
\label{fig:density_dist}
\end{figure}

The non-local dispersal analysis --- where dispersal kernels are fit to 22-year $\times$ site subsets of the data --- reveals a negative association between population density and dispersal (Fig. \ref{fig:density_dist}). This relationship applies to both short-distance and long-distance dispersal, respectively characterized by the median and $95^{\text{th}}$ percentile of predicted dispersal distances. A simple linear model, treating each fitted dispersal kernel as an observation, predicts that long-distance dispersal decreases from 10 km to 2 km across the data range. However, this negative relationship is noisy ($R^2 = 0.08, 0.19$ for the median and $95^{\text{th}}$ percentile respectively), even without accounting for conditional non-independence, i.e., the fact that successive years have similarly-shaped dispersal kernels. 

\begin{figure}[H]
\centering
\includegraphics[scale = 1]{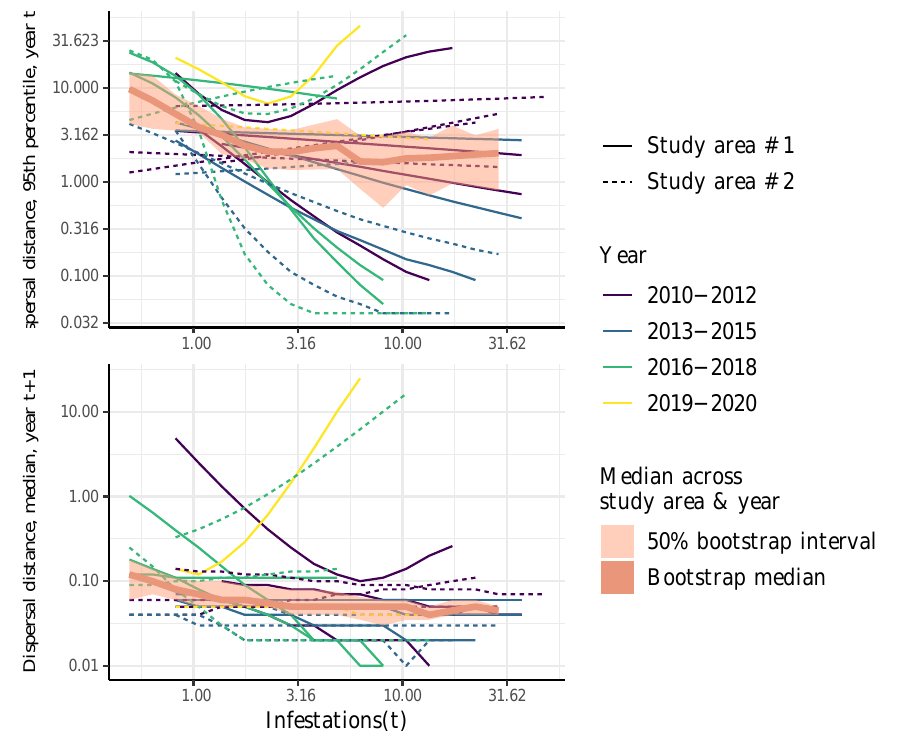}
\caption{The neighborhood-based dispersal model predicts a negative relationship between infestation density and dispersal distance, but the sign and strength of the relationship can vary substantially across space and time. The bootstrap intervals show the uncertainty in the spatiotemporal median of dispersal distance.}
\label{fig:density_dist_by_year}
\end{figure}

The neighborhood-based dispersal models --- where dispersal parameters depend on the density of nearby infestations --- also show an overall negative relationship between population density and dispersal distance (Fig. \ref{fig:density_dist_by_year}). However, the predominant pattern is one of spatiotemporal heterogeneity. There are substantial differences between years and study areas regarding the overall scale of dispersal. To a lesser extent, there is heterogeneity in the sign of the relationship between density and dispersal.

\begin{figure}[H]
\centering
\includegraphics[scale = 1]{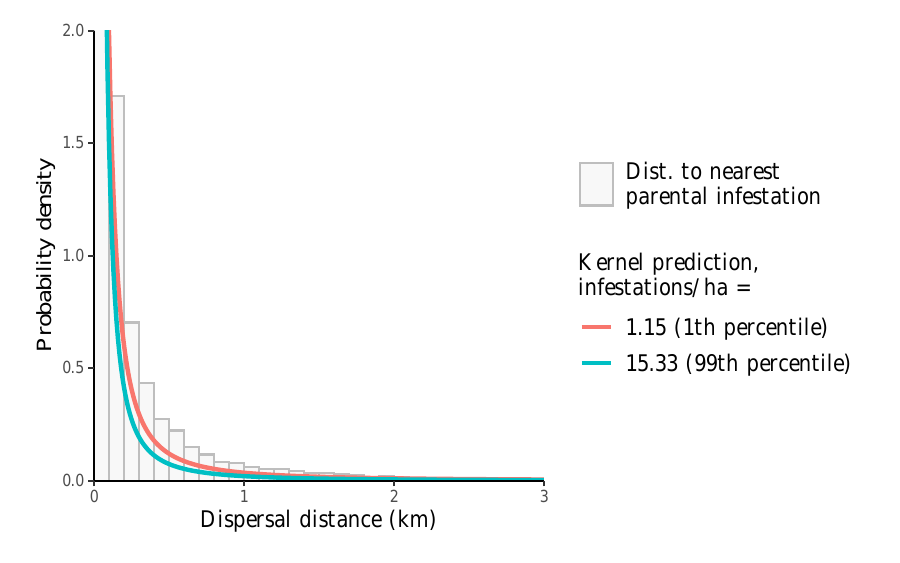}
\caption{The effect of infestation density on dispersal kernel shape. Depicted are \textit{marginal} dispersal kernels: $\int_{0}^{2\pi} D(r) r d\theta = 2\pi\,r\,D(r)$, where $r$ is the distance in kilometers, $\theta$ is the angle in radians, and $D(r)$ is the full density (\eqref{eq:student}). The histogram gives the frequency of distances between infestations and the closest ``parental infestation'', i.e., infestations in the previous year.}
\label{fig:DD_marginal_kernels}
\end{figure}

To cut through this heterogeneity, we can compute the median of the dispersal parameters ($\nu$ and $\rho$) across years and study areas for a range of neighborhood infestation densities. This provides a typical relationship between density and dispersal, which is small and negative (Fig. \ref{fig:DD_marginal_kernels}). To provide a more quantitative analysis, we may consider how dispersal changes as neighborhood infestation density increases from the 5th to the $95^{\text{th}}$ percentile (1.1 infestations/ha to 8.4 infestations/ha). The median dispersal distance decreases from 58 m to 42 m, whereas the $95^{\text{th}}$ percentile of dispersal distance decreases from 3.5 km to 1.6 km (Table \ref{tab:quantiles}). The difference in dispersal distance between low and high infestation densities increases (both in relative and absolute terms) for beetles that travel further from their origin. In other words, the effect of infestation density is more pronounced in the tails of the dispersal kernel.

\begin{table}[ht]
\centering
\begin{tabular}{llllllll}
  \hline
\multicolumn{2}{c}{\underline{Infestations/ha}} & \multicolumn{3}{c}{\underline{Median dispersal distance}} & \multicolumn{3}{c}{\underline{$95^{\text{th}}$ percentile dispersal distance}} \\ 
Quantile & Value & Median & $\text{BI}_{0.025\%}$ & $\text{BI}_{97.5\%}$ & Median & $\text{BI}_{0.025\%}$ & $\text{BI}_{97.5\%}$ \\ 
  \hline
0.100 &  1.149 & 0.058 & 0.036 & 1.029 & 3.398 & 1.465 &  9.703 \\ 
  0.250 &  1.533 & 0.055 & 0.031 & 0.522 & 2.648 & 1.177 &  8.504 \\ 
  0.500 &  1.916 & 0.054 & 0.030 & 0.442 & 2.080 & 0.804 &  7.463 \\ 
  0.750 &  3.448 & 0.051 & 0.024 & 1.332 & 1.967 & 0.307 &  8.569 \\ 
  0.900 &  6.130 & 0.044 & 0.015 & 5.982 & 1.619 & 0.064 & 15.309 \\ 
  0.950 &  8.429 & 0.042 & 0.012 & 5.487 & 1.572 & 0.043 & 15.937 \\ 
  0.975 & 11.111 & 0.043 & 0.015 & 9.249 & 1.686 & 0.043 & 20.600 \\ 
  0.990 & 15.326 & 0.041 & 0.015 & 0.078 & 1.766 & 0.068 & 12.235 \\ 
   \hline
\end{tabular}
\caption{Relationship between infestation density and typical dispersal distance. Quantiles are calculated across all neighborhoods, years, and study areas. Typical dispersal distance is the median across years and study areas where the given infestation density was observed (to avoid extrapolating beyond the data's range). Bootstrap intervals of the median used the were calculated with the same procedure. Empirical quantiles for probabilities 0.01, 0.025, and 0.05 were identical to the 0.1 probability (1.149 infestations/ha) and were thus omitted. } 
\label{tab:quantiles}
\end{table}

\section{Discussion} \label{Discussion}

We observe a moderate negative association between mountain pine beetle (MPB) population density and dispersal distance. While statistical models can be blunt instruments and are subject to various limitations (which we will discuss shortly), the observed negative relationship is likely real and not a statistical artifact. This conclusion is supported by a clear biological mechanism: higher population densities enable beetles to mass attack nearby healthy trees instead of seeking out weakened ones. This mechanism aligns with the observations of experienced field biologists. Anecdotally, only clusters of 3-6 (or more) infested trees are likely to produce nearby infestations in the following year (Katherine Bleiker, personal communication). Further reinforcing a negative relationship is the expectation of bias towards a positive relationship, stemming from the fact that we use infested tree density and their annual redistribution as proxies for beetle density and beetle dispersal. In the absence of any density-dependent dispersal behavior, more infestations mean more beetles; more beetles increase the probability of successful mass attacks, particularly far away from the parental infestation (i.e., in the tails of the dispersal kernel) where beetles are expected to be sparse. Thus, even in the absence of density-dependent dispersal, the average distance between parental and offspring infestations increases as a function of infestation density. 

Previous research on density-dependent dispersal provides a framework for interpreting our findings in the context of MPB biology. The literature shows that positive density-dependent dispersal is favored under specific conditions: when high-quality patch locations shift over time \citep{McPeek1992evolution, rodrigues2014evolution}, when dispersal costs are minimal \citep{Travis1999evolution, Poethke2002evolution}, or when local growth dynamics exhibit positive density dependence \citep{Travis1999evolution, rousset2012demographic}. MPB experiences a \textit{strong Allee effect}, an extreme form of positive density dependence where per capita growth rates become negative at sufficiently low population densities: negative growth occurs because there are not enough beetles to successfully mass attack a single tree \citep{safranyik1975interpretation, raffa1983role, boone2011efficacy, goodsman2017positive}. MPB's Allee effect means that patch quality is largely determined by the presence of conspecifics. Therefore, the locations of high-quality patches shift annually due to host tree depletion and the stochastic nature of MPB dispersal. Lastly, MPB dispersal is costly, with 40--80\% of emerging beetles dying during the dispersal phase \citep{safranyik1989mountain, latty2007pioneer, pope1980allocation}. This high mortality rate stems from several factors. MPB have limited survival time outside their host trees \citep{fuchs1985pre, reid2008fluorescent}, and prolonged exposure increases the chance of avian predation \citep{rust1930relation, stallcup1963method}. Further, beetles covering greater distances are less likely to engage in mass attacks, likely due to depleted energy reserves \citep{jones2019influence, jones2020mechanisms}. These observations collectively demonstrate the significant costs associated with MPB dispersal --- the selective basis for the observed negative relationship between population density and dispersal distance.

% Research on density-dependent dispersal should prioritize species-specific evidence, as meta-analyses indicate no consistent patterns in strength or direction across species. One study found that small-mammal research predominantly reports negative density-dependent dispersal \citep{rutherford2023go}, while another meta-analysis of mammal and bird studies mostly indicated positive density-dependent dispersal or ``no effect'' \citep{matthysen2005density}. Evidence of publication bias suggests negative density-dependent dispersal may be under-reported \citep{jreidini2024study}. The heterogeneity in observed density effects can be partially explained by varying study methodologies and the spatial scale of investigation \citep{morton2018dispersal, jreidini2024study}. Taken together, meta-analytic evidence suggests that it is inappropriate to presume a particular form of density-dependent dispersal. 

Although our study does not support a positive relationship between beetle density and long-distance dispersal, this possibility cannot be definitively ruled-out. Analysis of a dispersal model with quadratic effects, which could reveal a U-shaped relationship between density and dispersal, suggests this dynamic is not occurring in Alberta (Appendix \ref{quadratic}). However, a positive relationship might exist but only activate at higher beetle densities than those observed --- average infestation densities in Alberta are 2--3 orders of magnitude lower than in British Columbia. Heli-GPS from Alberta show less than 0.1 infestations/ha across the rocky mountain foothills, whereas the fixed-wing aerial survey data from British Columbia implies 10s to 100s of infestations/ha across vast areas in the Chilcotin Plateau. Unfortunately, Heli-GPS data for British Columbia during the peak outbreak years was never collected. An alternative test for density-dependent dispersal involves examining invasion speed, as positive density-dependent dispersal leads to acceleration of the invasion front over time \citep{dwyer2006resource, travis2009accelerating}. This analysis would be uninformative for two reasons: a limited sample size stemming from the short duration of MPB's range expansion across western Alberta (approximately 2005--2009), and the fact that acceleration in invasion speed could alternatively be attributed to MPB's Allee effect \citep{kot1996dispersal, wang2002integrodifference}.

% (Alan Carroll, personal communication)

Interestingly, the estimated effect of population density on dispersal distance is dwarfed by spatial and temporal heterogeneity in the typical level of dispersal. Visually, the shape of the dispersal kernel changes very little as population density changes (Fig. \ref{fig:DD_marginal_kernels}). We performed \textit{variance partitioning} on the logarithm of dispersal distance (Table \ref{tab:var_partition}). For both the median and $95^{\text{th}}$ percentile of distance (representing short and long-distance dispersal respectively), we found that spatiotemporal heterogeneity accounted for 82\% and 75\% of the total variance. This heterogeneity is evident in the erratic eastward expansion of MPB across Alberta, varying from 20 to 220 km per year \citep{cooke2017predicting}. There is no shortage of factors that could plausibly explain this variability: the occurrence of wind storms during the beetles' 1--4 week flight period; adverse weather conditions during the flight period  (e.g., too much wind or precipitation, \citealp{gray1972emergence}; \citealp{musso2023pine}, Ch. 2); cold winter temperatures affecting development and subsequent flight capability \citep{nijholt1967moisture, shegelski2019morphological}; tree defensive compounds --- thought to vary over large spatial scales \citealp{cudmore2009geographic, clark2012legacy} --- affecting development and subsequent flight capability \citep{manning2013sub}; and volatiles from deciduous non-host trees affecting flight propensity \citep{jones2021effect}.

Our statistical models have several limitations. 1) A spurious negative correlation between beetle density and dispersal distance might arise from the confounding effect of site quality: beetles are abundant in prime locations --- dense stands of large trees --- and do not need to disperse far to find suitable hosts. We do not believe that this is a major influence on our results, since we intentionally selected our study areas to contain contiguous lodgepole pine forest. 2) Our dispersal kernels only extend for 5 km in each direction (see \textit{Methods}). 3) Our models assume that the location of every infestation relative to the source is an i.i.d. random variable given by the dispersal kernel, whereas in reality, MPB infestations are more spatially autocorrelated than expected (after accounting for local dispersal), due to aggregation behavior and non-isometric dispersal patterns. Consequently, the standard errors for the dispersal hyperparameters ($\alpha_0, \alpha_1, \alpha_2, \beta_0, \beta_1, \& \beta_2$) are smaller than they should be, although the large amount of data suggests a high degree of precision. The standard solution, a spatial Gaussian process, is computationally infeasible with our large dataset. Therefore, instead of reporting and emphasizing the standard errors of dispersal hyperparameters, we acknowledge uncertainty by focusing on between-year and between-site heterogeneity (e.g., Fig. \ref{fig:density_dist_by_year}).

%Despite the negative relationship observed between MPB density and dispersal range, focusing control measures in central Alberta may still be strategically important. As discussed in the \textit{Introduction}, the most significant risk of further eastward range expansion is a long-distance dispersal event from Central Alberta to western Saskatchewan. But even if population density doesn't modulate the probability of long-distance dispersal, it likely modulates the probability that beetles become established in western Saskatchewan. If a source population is small, then the few beetles dispersing long distances will encounter densities below the Allee threshold --- the population density at which per capita growth rates are zero --- at their destination. This phenomenon occurs in the North American gypsy moth invasion, where cyclical population dynamics and an Allee effect drive quasi-periodic range expansion ``pulses'' \citep{johnson2006allee}. Therefore, even if control measures can't completely suppress incipient outbreaks of MPB, control can still be justified as a means of preventing range expansion.

Future research on MPB dispersal should focus on explaining why dispersal distances vary across time and space. We suggest starting by investigating how wind speed and direction during the MPB's flight period affect dispersal, given that windstorms have clearly facilitated MPB's journey across the Canadian Rocky Mountains \citep{jackson2008radar, chen2017climatic}, and that wind direction is highly predictive of spruce budworm dispersal \citep{sturtevant2013long}. Our study challenges the prevailing notion of a positive relationship between MPB density and long-distance dispersal, instead providing evidence for a small negative association. This result underscores the complexity of MPB dispersal, the value of high-quality (Heli-GPS) data, and the essential role of simple statistical models in understanding ecological phenomena.

\pagebreak

\section{Acknowledgements}
The authors would like to thank K{'e}van Rastello, Micah Brush, and Leanne Petro for their feedback and relevant conversations. Funding for this research has been provided through grants to the TRIA-FoR Project to ML from Genome Canada (Project No. 18202) and the Government of Alberta through Genome Alberta (Grant No.~L20TF).

\section{Data sources} \label{Data and code availability statement}
Code and data will be made available on Zenodo (\url{TBD}).

\section{Author contributions}
Evan C. Johnson conceived of the project, performed the analysis, and wrote the first draft; both authors contributed critically to the drafts and gave final approval for publication.

\begin{appendices}
\counterwithin{figure}{section}
\counterwithin{table}{section}

\section{Additional figures and tables}

\begin{figure}[H]
\centering
\includegraphics[scale = 1]{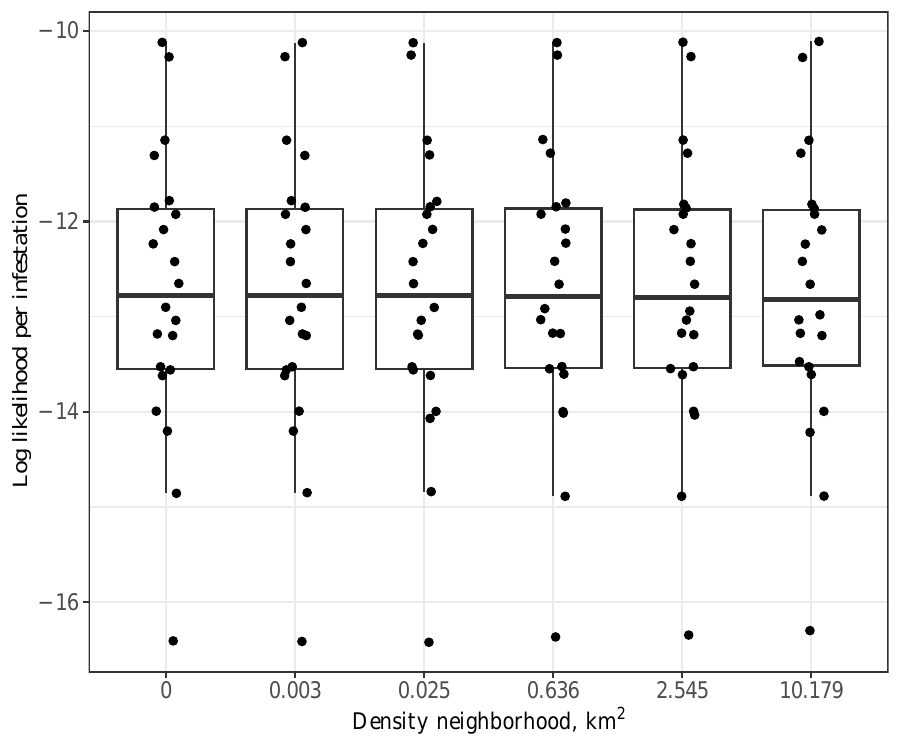}
\caption{There is no clear relationship between model fit and the size of the neighborhood in the \textit{neighborhood-based dispersal models}. Each point is a distinct year $\times$ study area combination.}
\label{fig:DD_dens_boxplot}
\end{figure}

\begin{figure}[H]
\centering
\includegraphics[scale = 1]{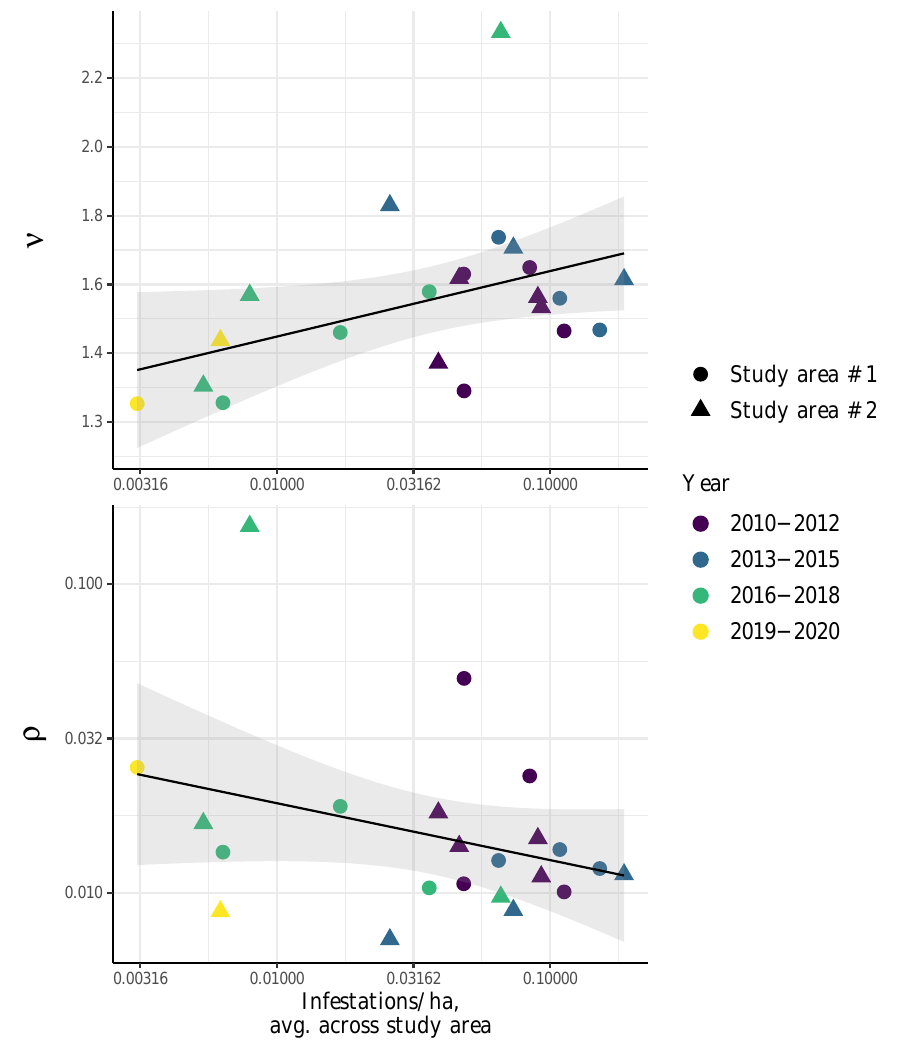}
\caption{TBD}
\label{fig:density_params}
\end{figure}

% latex table generated in R 4.4.0 by xtable 1.8-4 package
% Tue Jun 25 13:06:57 2024
\begin{table}[H]
\centering
\begin{tabular}{lll}
  \hline
 & \multicolumn{2}{c}{Variation in ...} \\ \cline{2-3}

Source of variation & Log(median dist.) & Log(95th perc. dist.) \\ 
  \hline
\makecell{Across infestation densities,\\ within areas \& years} & 0.15 & 0.37 \\ 
  Between areas \& years & 0.66 & 1.00 \\ 
   \hline
\end{tabular}
\caption{Heterogeneity across study areas and years accounts for most of the variation in dispersal distance: 81\% of the variation in the logarithm of the median dispersal distance, and 73\% of the variation in the logarithm of the $95^{\text{th}}$ percentile of dispersal distance. } 
\label{tab:var_partition}
\end{table}

% \begin{figure}[H]
% \centering
% \includegraphics[scale = 1]{figures/distance_time_series.pdf}
% \caption{TBD}
% \label{fig:distance_time_series}
% \end{figure}

% \begin{figure}[H]
% \centering
% \includegraphics[scale = 1]{figures/infest_time_series.pdf}
% \caption{TBD}
% \label{fig:infest_time_series}
% \end{figure}

\section{Quadratic neighborhood models} \label{quadratic}

In the Introduction, we proposed two distinct mechanisms that describe how beetle population density might influence dispersal distance. At low population densities, an increase in density may decrease dispersal distance. The proposed mechanism is that a higher number of beetles increases the probability of overcoming the resin defenses of nearby infested trees, thus reducing the need to search out a tree that is suitably large and recently damaged (for instance, by root rot or wind damage). Conversely, at high population densities, further increases in density may lead to an increase in dispersal distance. The proposed mechanism is a higher concentration of aggregation pheromones produced by the beetles, which become repellent at sufficiently high concentrations, prompting beetles to fly above the canopy where they may be taken by wind.

These two mechanisms, which operate in different density regimes, together suggest a U-shaped relationship between beetle density and dispersal distance: at low densities, increasing density decreases dispersal distance, while at high densities, further increases in density result in longer dispersal distances.

The linear model presented in the main text is insufficient to capture this potential U-shaped pattern. Therefore, we have expanded our model to include quadratic terms. This enhanced model describes the dispersal parameters ($\rho$ \& $\nu$) as functions of six hyperparameters: $\alpha_0, \alpha_1, \alpha_2, \beta_0, \beta_1, \& \beta_2$. The equations, previously \eqref{eq:rho} \& \eqref{eq:nu} in the main text, become

\begin{equation}
\rho_{t}(x) = \exp\left[\alpha_0 + \alpha_1 \log\left(\sum_{y \in N(x)} I_{t-1}^*(y) \right)\right],
\end{equation}

\begin{equation}
\nu_{t}(x) = \exp\left[\beta_0 + \beta_1 \log\left(\sum_{y \in N(x)} I_{t-1}^*(y) \right)\right] + 1.
\end{equation}

\begin{figure}[H]
\centering
\includegraphics[scale = 1]{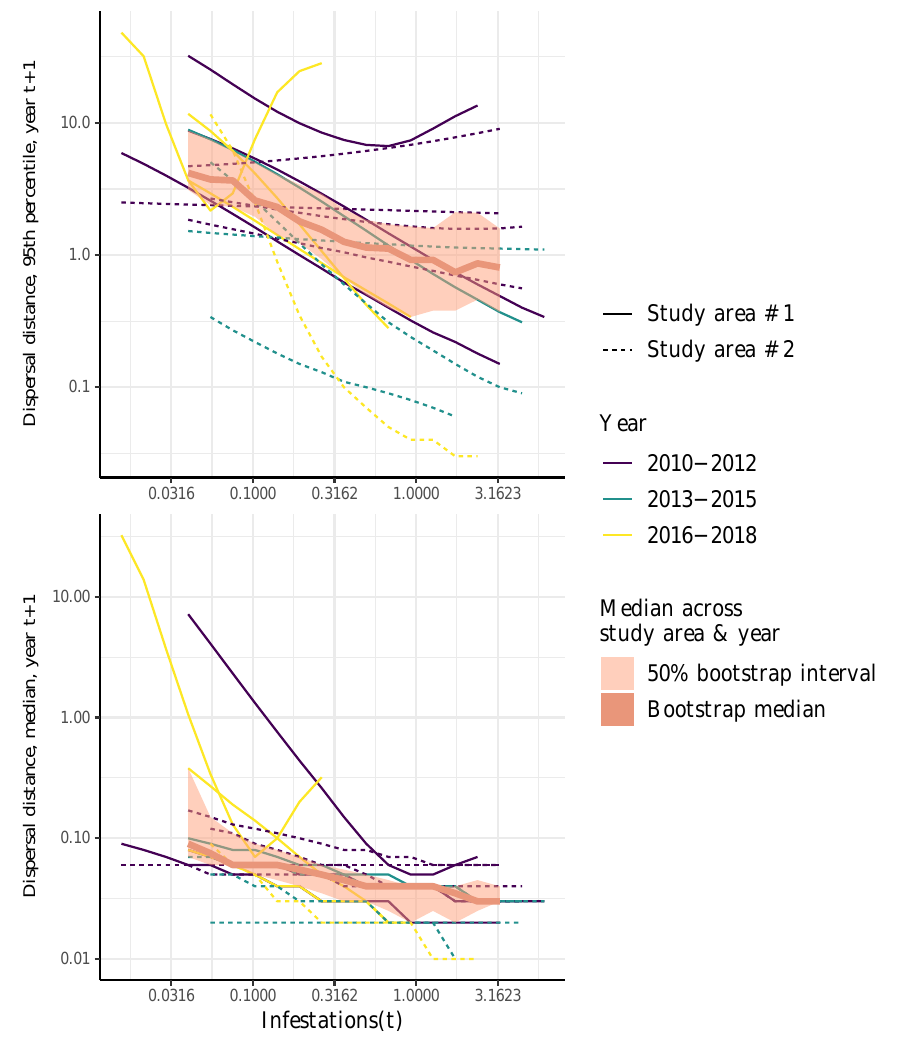}
\caption{The neighborhood-based dispersal model with quadratic effects of infestation density still predict an overall negative relationship between infestation density and  dispersal distance.}
\label{fig:density_dist_by_year_quad}
\end{figure}

By incorporating these quadratic terms, the expanded model allows for a more accurate representation of the relationship between beetle density and dispersal distance. However, the results produced by this expanded model were qualitatively identical to those of the simpler model (Fig. \ref{fig:density_dist_by_year_quad}). For the sake of simplicity in the presentation, and because the Nelder-Mead algorithm failed to converge for a larger fraction of the year-site subsets of the data, we decided to present the simpler model in the main text.

\end{appendices}
% \bibliography{\Sexpr{here("scripts/R/writeup/refs_for_spatiotemporal_coexistence_paper.bib")}}
%\bibliographystyle{apalike}
%\bibliographystyle{plain}
\bibliographystyle{apalike}
\bibliography{mpb_refs.bib}

% \section{References}
% \printbibliography[heading=bibintoc]

\end{document}